# Quantum longitudinal and Hall transport at the $LaAlO_3/SrTiO_3$ interface at low electron densities

Yanwu Xie,[1,2,*] Christopher Bell,[2] Minu Kim,[2] Hisashi Inoue,[1] Yasuyuki Hikita,[2] Harold Y. Hwang[1,2,*]

[1]*Geballe Laboratory for Advanced Materials, Department of Applied Physics, Stanford University, Stanford, California 94305, USA*
[2]*Stanford Institute for Materials and Energy Sciences, SLAC National Accelerator Laboratory, Menlo Park, California 94025, USA*

*Emails: xieyanwu@stanford.edu, hyhwang@stanford.edu

We examined the magneto-transport behavior of electrons confined at the conducting $LaAlO_3/SrTiO_3$ interface in the low sheet carrier density regime. We observed well resolved Shubnikov-de Haas quantum oscillations in the longitudinal resistance, and a plateau-like structure in the Hall conductivity. The Landau indices of the plateaus in the Hall conductivity data show spacing close to 4, in units of the quantum of conductance. These experimental features can be explained by a magnetic breakdown transition, which quantitatively explains the area, structure, and degeneracy of the measured Fermi surface.

Recently, there has been significant interest in quantum transport [1–7] and the electronic structure [8–14] of quasi-two-dimensional electron gases (q2DEGs) generated in $SrTiO_3$, in which a variety of fascinating properties have been observed [15,16]. However, there are currently few measurements which clearly resolve the nature of the electronic band structure of these q2DEGs. Although Shubnikov-de Haas (SdH) quantum oscillations, which directly measure the Fermi surface, have been observed [1–7,17], they were often poorly resolved. In particular, the Fermi surface constructed from these measurements only accounts for a small fraction of the carriers measured by the Hall effect.

Compared with many conventional semiconductors, $SrTiO_3$ has carriers of much larger effective mass, $m^*$, which results in a significantly smaller separation of Landau levels in the presence of a magnetic flux $B$, $heB/2\pi m^*$ ($h$ is the Planck constant and $e$ is the electronic charge). Therefore, it is correspondingly more challenging to realize truly 2D quantum transport in $SrTiO_3$. A variety of techniques have been utilized to approach the 2D limit in $SrTiO_3$, including delta-doping [1,4,7], electrostatic field-effect doping [18–22] and heterostructuring [6,15,16,23]. In the latter category the q2DEG formed between $TiO_2$-terminated {100} $SrTiO_3$ and $LaAlO_3$ is intensively studied [15,16]. It has been observed that $LaAlO_3/SrTiO_3$ heterostructures grown at lower temperature give relatively lower sheet carrier density and higher mobility [2]. A continuous increase in mobility with decreasing carrier density have been demonstrated when $LaAlO_3/SrTiO_3$ heterostructures were tuned by $LaAlO_3$ surface treatments such as charges and adsorbates [24]. These studies suggest promising avenues to explore quantum transport in the low electron density regime, possessing simultaneously favorably high mobility and reduced subband complexity.

In this work, we studied the magneto-transport behavior of the q2DEG confined at the $LaAlO_3/SrTiO_3$ interface in the low sheet carrier density regime. Our results reveal two compelling features: 1) the presence of plateaus in the Hall conductivity that can be assigned to Landau filling indices with an interval close to 4, implying a 4-fold degeneracy in the band structure; 2) a transition of the frequency of the SdH quantum oscillations (from $f$ to ~3$f$) with increasing magnetic field. These features can be understood by considering magnetic breakdown orbits and account for all of the carriers.

We use $LaAlO_3/SrTiO_3$ heterostructures of lower sheet carrier density and higher mobility than in previous studies [2,3,5]. The samples were prepared by depositing $LaAlO_3$ thin films on $TiO_2$-terminated {100} $SrTiO_3$ substrates by pulsed laser deposition, at 650 ˚C, under $1.3 \times 10^{-3}$ Pa of $O_2$ [2]. The samples were further treated by writing surface charge using a biased conductive atomic force microscope (CAFM) probe [24] or using a capping layer [5] to produce Hall mobilities up to 10,000 $cm^2V^{-1}s^{-1}$. Here we present transport measurements of four different samples. The details of the heterostructures and surface processing are summarized in Table 1. The samples B & C were treated by CAFM under a bias of -8 V, at ambient environment with a relative humidity of ~35 %. The sample D was capped with a 140 nm amorphous $LaAlO_3$ (a-LAO) layer deposited at room temperature. All samples were measured using a standard Hall bar configuration with a current in a range of 0.2-1 $\mu$A. The electrical contacts to the buried q2DEG were made by ultrasonic bonding with Al wires. The samples A, C & D were measured in a $^4$He cryostat with a $^3$He insert. The sample B was measured in a dilution refrigerator with a base temperature of 10 mK. All of these samples showed metallic conduction down to low temperatures [Fig. S1].

In Fig. 1 we show the raw longitudinal and Hall resistances, $R_{xx}$ and $R_{xy}$, as a function of $B$ for the four samples, measured at $T \leq 0.5$ K. Significant SdH oscillations are observed in all cases. In the samples A & C [Figs. 1(a) & 1(c)] the amplitude of oscillations in $R_{xx}$ at $B > 9$ T is more than 50 % of the total resistance, indicating the high quality of the samples. $R_{xy}(B)$ is close to linear except for the superimposed plateau-like structure (the slight non-linearity for sample B may originate from parallel conduction channels). From the low-field ($B < 5$ T) Hall coefficient $R_H$, the sheet carrier density $n_{Hall} = -1/eR_H$ and Hall mobility $\mu = R_H/R_{xx}(0)$ were evaluated to be in the range of 4.7 to 7.5 × $10^{12}$ $cm^{-2}$ and 5,500 to 10,000 $cm^2V^{-1}s^{-1}$, respectively [Table 1].

As shown in Fig. 1, the plateau-like features in $R_{xy}(B)$ become more significant with increasing $B$. In the inset of each panel we plot the Hall conductivity, $G_{xy} = R_{xy}/(R_{xx}^2 + R_{xy}^2)$, versus $B$ in the high field range. At first glance these plateaus in $G_{xy}$ can be approximately assigned to the quantum filling indices with an interval close to 4. To better understand the indices of these plateaus, in Fig. 2 we plot the high field $G_{xy}(B)$ and its derivative in the same figure for each sample, and examine the corresponding values of $G_{xy}$ at the local maxima of $dG_{xy}/dB$. We found that the plateau positions are not precisely at integer multiples of $e^2/h$ and the indexing shows both odd and even values in different samples, suggesting that we are not in a purely single-band quantum Hall regime. These observations might be related to the fact that more than one effective mass carriers are involved in the quantization, as discussed below. Nonetheless the ~$4e^2/h$ spacing of $G_{xy}$ implies an apparent degeneracy in the electronic structure of the q2DEG.

Next we analyze the quantum oscillations in $R_{xx}(B)$ and $R_{xy}(B)$ in more detail. In the following we focus on the data from the sample A, but similar results are found for the other samples. To clearly resolve the oscillations from the non-oscillatory background we performed second and first order differentiation of $R_{xx}(B)$ and $R_{xy}(B)$ respectively, using the data shown in Fig. 1(a) [see also Fig. S3]. Figure 3(a) shows $dR_{xx}^2/dB^2$ and $dR_{xy}/dB$ versus $1/B$. Two distinct sets of oscillations in $dR_{xx}^2/dB^2$ are found, with a crossover in behavior around ~4.3 T ($1/B$ = 0.23 $T^{-1}$). A Fourier transform (FT) of these $dR_{xx}^2/dB^2$ data gives two frequencies, $f_{xx1}$ ~ 20 T and $f_{xx2}$ ~ 60 T [Fig. 3(b)]. The inverse FT shows that the oscillations in low field range ($B$ < 4.3 T) correspond to $f_{xx1}$ and the oscillations in high field range ($B$ > 4.3 T) correspond to the superposition of $f_{xx1}$ and the dominant $f_{xx2}$ [Fig. S5].

Further insight is obtained by performing a standard fan diagram analysis of the oscillations, where we plot the index of the $dR_{xx}^2/dB^2$ extrema versus $1/B$. As shown in Fig. 3(c), the magnitudes of the slope of the fan diagram in the low and high field regimes, $S_L$ and $S_H$ respectively, agree well with $f_{xx1}$ and $f_{xx2}$. The crossover of $S_L$ and $S_H$ corresponds to the transition of the two sets of oscillations. Here in the high field range of the fan diagram analysis only the oscillations which correspond to $f_{xx2}$ have been distinguished. Analysis of the oscillations in $dR_{xy}/dB$ gives similar results [Figs. 3(a), 3(b) & 3(c)], although the absolute values of the frequencies and slopes are slightly smaller. This difference is ascribed to the relatively lower resolution of the oscillations in $dR_{xy}/dB$ compared to those in $dR_{xx}^2/dB^2$.

From the temperature dependence of the amplitude of the oscillations in $R_{xx}$ [Fig. S4(a)] we obtain a cyclotron effective mass $m^*$= 0.9 ± 0.1 $m_e$ at $B \approx 4$ T and $m^*$= 1.30 ± 0.04 $m_e$ at $B \approx 11$ T [Table 1 & Fig. S6]. The estimated Dingle temperature, $T_D$, is ~1.3 (~ 2.0) K in the high- (low-) field range, smaller than previous reports in similar q2DEGs [1,2], indicating relatively less broadening of the Landau levels in the present samples.

We now discuss the possible electronic structures which can explain the ~ 4-fold degeneracy, the $S_H/S_L$ ratio, and the effective mass variation. First of all, we can exclude Zeeman spin splitting as the source of the field-dependent frequency change of the SdH oscillations, since it would give $S_H/S_L = 2$, rather than 3. The conduction band of bulk $SrTiO_3$ consists of a light band and a heavy band which is relatively close in energy, as well as a split-off band that is expected to exceed the Fermi level in these samples of low carrier density [9,25–28]. Quantum confinement at the interface produces a series of 2D subbands [9,11,12] derived from the light and heavy bands, forming the q2DEG. We consider a case where one light and one heavy subband are occupied, as suggested theoretically for a q2DEG of comparable carrier density with our present samples [9]. Figure 4(a) shows the schematic Fermi surface, in which the inner circle and the outer star represent the light and the heavy subbands respectively. In this scenario it is natural to assign $S_L$ and $m^*$(low-$B$) to the circle and $S_H$ and $m^*$(high-$B$) to the one of two possibilities: the star or magnetic breakdown (MB) orbits between the star and the circle [26]. The former can be ruled out since it predicts no degeneracy.

In Fig. 4(a) we sketched two representative MB orbits, MB1 (black and shaded) and MB2 (yellow dashed). We note that MB1 is the lowest order closed breakdown orbit (2nd) with smallest area and effective mass, and would be the first expected to be observed. Although it is possible to construct other MB orbits, they are not responsible for the present observations, being ruled out by a combination of their inappropriate degeneracy, $m^*$, or enclosed area, following application of the same discussions as below. We assume a Landé $g$-factor of ~2, as has been observed in bulk $SrTiO_3$ [29], and in weak-localization studies of the $LaAlO_3/SrTiO_3$ interface [13]. In this case, using $m^* = 1.3\ m_e$, the Zeeman spin splitting ($g\mu_B B$, where $\mu_B$ is the Bohr magneton) is comparable with the Landau splitting ($heB/2\pi m^*$): 1.3 meV *vs* 0.9 meV at $B$ = 10 T. Hence, the intrinsic spin degeneracy of 2 should be lifted at all fields [7]. However, the spin-split up (down) Landau levels may overlap with the adjacent down (up) levels when the splitting energies of the two effects are close to one another, causing apparent spin degeneracy that cannot be resolved due to the finite energy resolution of the SdH oscillation measurements (see discussion in Supplemental Material, Section E). Hence the apparent 4-fold degeneracy observed in $G_{xy}(B)$ can be explained by either, a) the 4 equivalent MB1 orbits without spin degeneracy or b) the 2 equivalent MB2 orbits with effective spin degeneracy.

From the Onsager relation the frequency of the SdH oscillations (or the slope in the fan diagram) is proportional to the enclosed area of the electronic orbit at the Fermi surface, $A$. The relation between the two dimensional carrier density, $n_{2D}$, and $A$ is, $A = 4\pi^2 \times n_{2D}$/degeneracy. With the light electron density $n_l = eS_L/h$ (no spin degeneracy case) and heavy electron density $n_h = n_{Hall} - n_l$, we calculated the $A$ of light and heavy subbands, and thus the $A$ of MB orbit, based on the geometrical relation schemed in Fig. 4 (a). We found that in the case a) $A$(MB1) is ~3$A$(circle), while in the case b), with presumed 2-fold spin degeneracy, $A_F$(MB2) is ~2.4$A$(circle). Thus the case a) matches the observed $S_H/S_L$ (or $f_{xx2}/f_{xx1}$) ratio of 3. In addition, from the definition of the cyclotron effective mass [30,31], $m^* = \frac{h^2}{8®^3}\frac{\partial A}{\partial E}|_{E=E_F}$, and the geometrical relations shown in Fig. 4(a), we have $m^*(\mathrm{MB1}) \cong \frac{1}{4}m_h^* + \frac{3}{4}m_l^*$ and $m^*(\mathrm{MB\,2}) \cong \frac{1}{2}m_h^* + \frac{1}{2}m_l^*$. Substituting $m^*$(MB) and $m_l^*$ with the above measured values, we get $m_h^*$ = 2.5 ± 0.5 $m_e$ and 1.7 ± 0.2 $m_e$ in the cases a) and b), respectively. Again, the $m_h^*$ deduced in the case a) is in better agreement with the values given by previous theoretical calculations [25] and experimental measurements [11]. Therefore we conclude that the case a) is the scenario most consistent with the experimental observations. Figure 4(b) shows the schematic energy band structure based on the case a). Assuming a parabolic dispersion relation, the Fermi energy $E_F$ is calculated to be ~8.0 meV, and the separation between the light and the heavy subbands is ~5.5 meV at the $\Gamma$ point. We note that for the occurrence of magnetic breakdown the inequality $heB/2\pi m^* \geq E_g^2/E_F$ must be satisfied, where $E_g$ is the energy difference between the two subbands at the tunneling point [26,32]. This criterion suggests that the heavy subband should have a steeper dispersion relation in the $k_{MB}$ direction than the light subband.

In summary, we have demonstrated well-resolved SdH quantum oscillations and Hall plateaus in the q2DEG confined at the $LaAlO_3/SrTiO_3$ interface in the low density regime. We found that the indices of the Hall plateaus have an interval of ~4, and that the high- and low-field frequencies of the SdH oscillations have a ratio of ~3. In a magnetic breakdown scenario, this 4-fold degeneracy quantitatively accounts for all the mobile carriers in the observed quantum transport.

We are grateful to N. P. Ong, S. Raghu, G. Xu and H. Yao for helpful discussions. This work is supported by the Department of Energy, Office of Basic Energy Sciences, under Contract No. DE-AC02-76SF00515.

**Supplemental Material** accompanies this paper.

**Figure legends**

**FIG. 1.** Quantum oscillations in magnetoresistance. $R_{xx}$ and $R_{xy}$ as a function of $B$ in (a) sample A, (b) sample B, (c) sample C, and (d) sample D. The dashed lines are linear fits to the low field ($B$ < 5 T) $R_{xy}$ data. In the inset of each panel the Hall conductance, $G_{xy}$, as a function of $B$ is shown in unit of $e^2/h$. The indices label the integer positions of $e^2/h$. The lower inset of (d) shows a photograph of a typical sample.

**FIG. 2.** Positions of plateaus in $G_{xy}$. The corresponding $G_{xy}$ (in unit of $e^2/h$) at local maxima in $dG_{xy}/d$B data of (a) sample A, (b) sample B, (c) sample C, and (d) sample D. The numbers in the parentheses indicate the closest integer indices, as labeled in the insets of Fig. 1.

**FIG. 3.** Fourier transform (FT) and fan diagram analysis of quantum oscillations. The oscillations are extracted from the magnetoresistance data of sample A shown in Fig. 1(a). (a) $\mathrm{d}R_{xx}^2/\mathrm{d}B^2$ and $\mathrm{d}R_{xy}/\mathrm{d}B$ as a function of $1/B$. (b) FT of the data shown in (a). (c) Index of extrema (integer $n$ for maximum and $n$ + 1/2 for minimum) in $\mathrm{d}R_{xx}^2/\mathrm{d}B^2$ as a function of $1/B$. (d) Index of extrema in $\mathrm{d}R_{xy}/\mathrm{d}B$ (integer $n$ for maximum and $n$ + 1/2 for minimum.) as a function of $1/B$. In (c) and (d) the indices are chosen to make the intercept of $S_H$ between 0 and 1.

**FIG. 4.** Schematic electronic orbits. (a) Fermi surface when the q2DEG consists of one light ($l$) and one heavy ($h$) subband, showing the inner light circle and the outer heavier star-shaped geometry. The dark and shaded MB1 and the yellow dashed MB2 indicate two possible magnetic breakdown (MB) orbits. The green dots indicate the MB tunneling paths. By symmetry there are 4 equivalent MB1 orbits and 2 for MB2. (b) Energy *vs* momentum dispersion along the $k_x$ and $k_{MB}$ directions for the $l$ and $h$ subbands.

**Table 1. Structure and electronic properties of samples.** The width ($W$) of Hall bar and the thickness of $LaAlO_3$ film ($t_{LAO}$). In the calculation of electron density, $S_H$ and $S_L$ are from the analysis of $dR_{xx}^2/dB^2$; the corresponding values from the analysis of $dR_{xy}/dB$ are shown in the parenthesis.

| | | | | Electron density ($\times 10^{12}$ cm$^{-2}$) | | | | Effective mass $m^*$ ($m_e$) | |
|---|---|---|---|---|---|---|---|---|---|
| **Sample** | $W$ (μm) | $t_{LAO}$ (uc) | **Surface treatment** | $n_{Hall}$ | $eS_H/h$ | $eS_L/h$ | $\mu$ (cm$^2$V$^{-1}$s$^{-1}$) | ~11 T | ~4 T |
| **A** | 10 | 8 | None | 4.7 | 1.45 (1.24) | 0.48 (0.45) | 7,600 | 1.30 ± 0.04 | 0.9 ± 0.1 |
| **B** | 5 | 10 | CAFM | 7.5 | 1.59 (1.64) | 0.58 (0.58) | 7,100 | 1.32 ± 0.03 | - |
| **C** | 5 | 5 | CAFM | 5.1 | 1.44 (1.38) | 0.52 (-) | 5,500 | 1.19 ± 0.03 | 0.72 ± 0.04 |
| **D** | 10 | 10 | Cap 140 nm a-LAO | 4.8 | 1.28 (1.23) | 0.55 (-) | 10,000 | - | - |

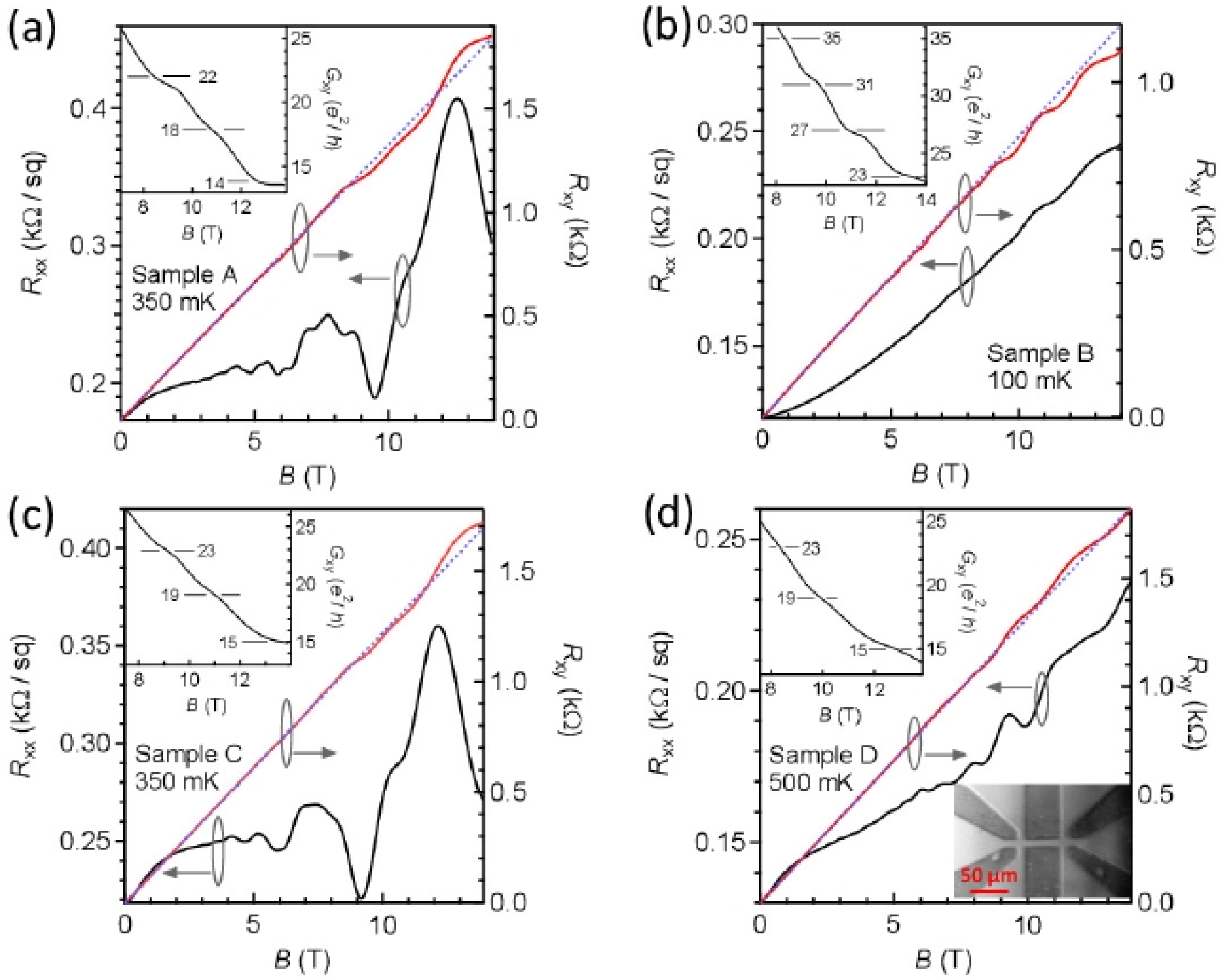

FIG. 1, Y. Xie et al.

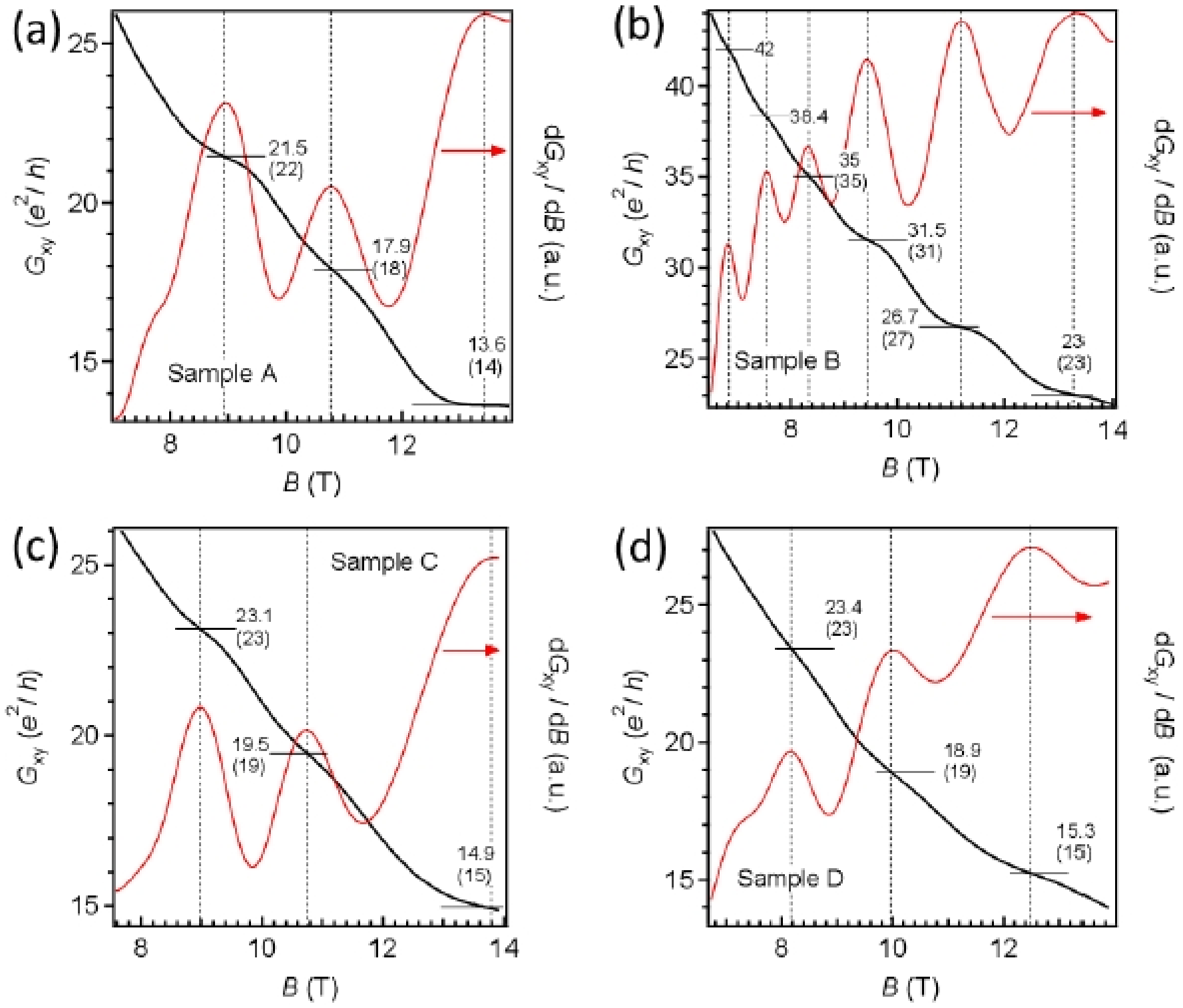

FIG.2, Y. Xie et al.

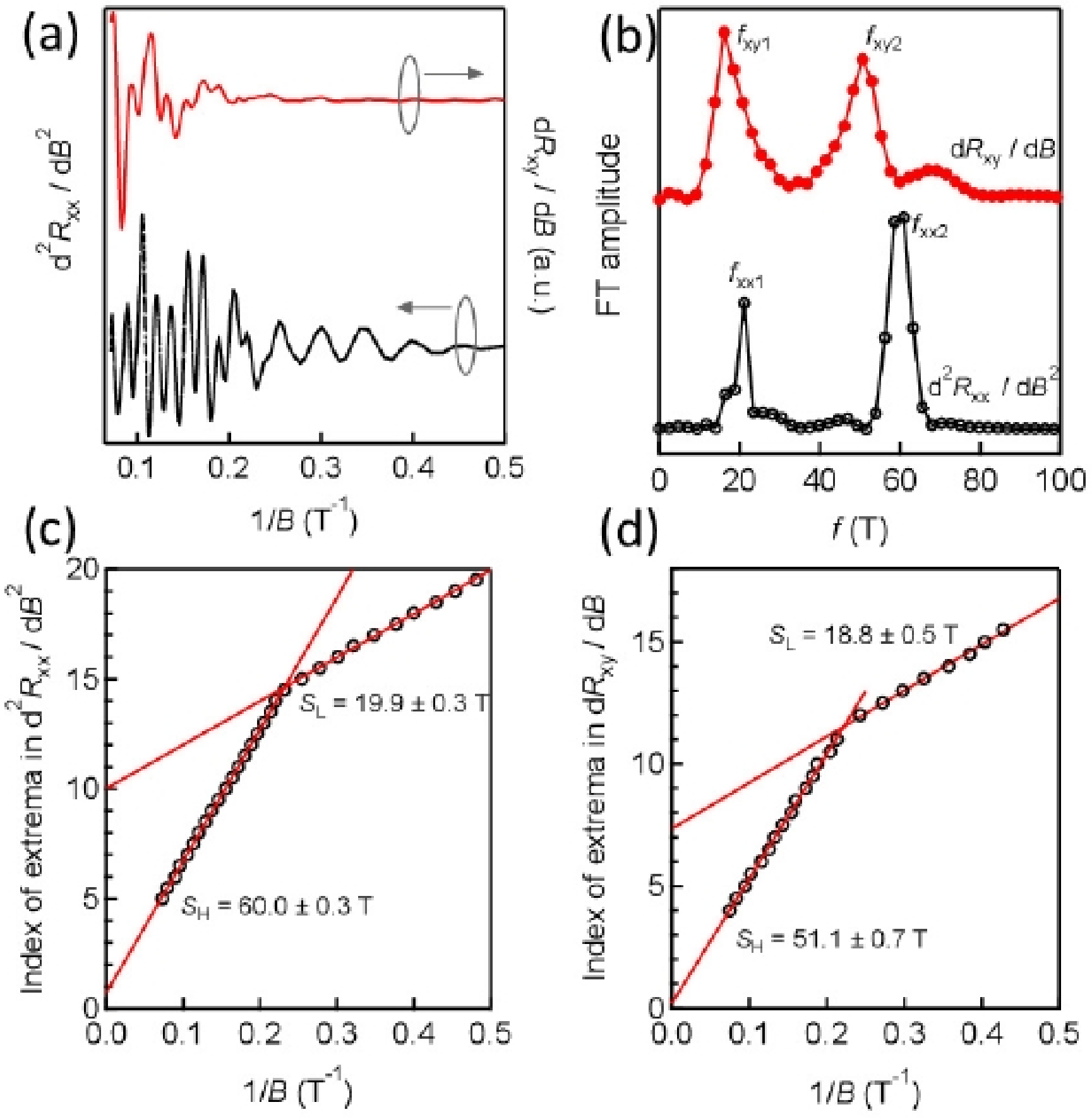

FIG. 3, Y. Xie et al.

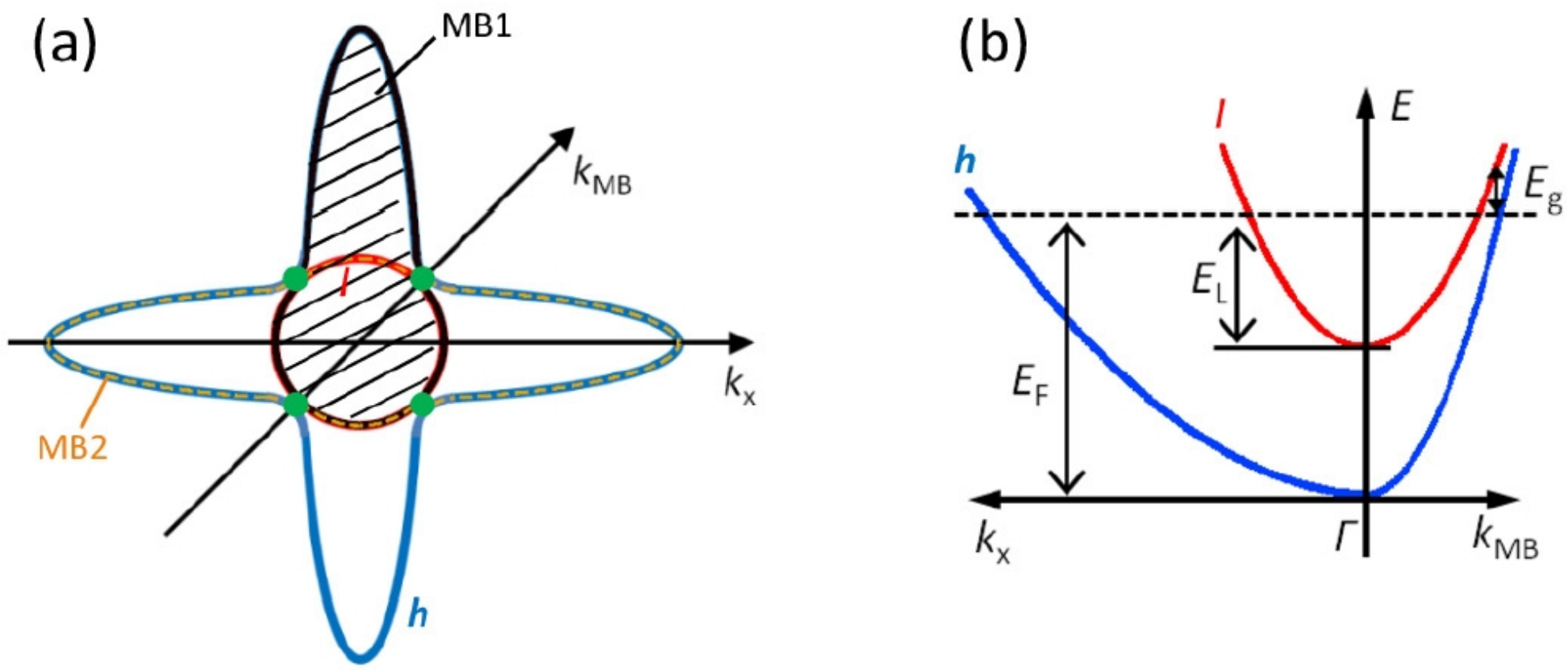

FIG.4, Y. Xie et al.

# Supplemental Material for "**Quantum longitudinal and Hall transport at the $LaAlO_3/SrTiO_3$ interface at low electron densities**"

Yanwu Xie,[1,2,*] Christopher Bell,[2] Minu Kim,[2] Hisashi Inoue,[1] Yasuyuki Hikita,[2] Harold Y. Hwang[1,2,*]

[1]*Geballe Laboratory for Advanced Materials, Department of Applied Physics, Stanford University, Stanford, California 94305, USA*
[2]*Stanford Institute for Materials and Energy Sciences, SLAC National Accelerator Laboratory, Menlo Park, California 94025, USA*

*Emails: xieyanwu@stanford.edu, hyhwang@stanford.edu

**Contents**

**SM A. Materials and Methods**

**1. Sample growth.** The samples were prepared by depositing $LaAlO_3$ thin films on $TiO_2$-terminated {100} $SrTiO_3$ substrates by pulsed laser deposition. The $SrTiO_3$ substrates were patterned into six-probe Hall bars [inset of Fig. 1(d) in the main text] by conventional optical lithography and lift off of an amorphous $AlO_x$ hard mask [1,2]. Before growth, the substrates were pre-annealed at 950 ˚C in $6.7 \times 10^{-4}$ Pa of $O_2$ for 30 minutes. The growth of $LaAlO_3$ was performed at 650 ˚C, under $1.3 \times 10^{-3}$ Pa of $O_2$. After growth, the samples were in-situ post-annealed at 550˚C, in $2\times10^4$ Pa of $O_2$ for 1 hour. The laser fluence at the $LaAlO_3$ single-crystal target was 0.7 $Jcm^{-2}$. The thickness of $LaAlO_3$ films was monitored using in-situ reflection high-energy electron diffraction.

**2. Capping layer.** For sample D, in addition to the normal growth processes, we deposited a 140 nm amorphous $LaAlO_3$ (a-LAO) cap layer at room temperature, under 10 Pa of $O_2$, and followed by a 3-hour anneal at 300 ˚C in a flow of $O_2$ at a pressure of $10^5$ Pa.

**3. Conductive atomic force microscopy (CAFM) treatment**. As described elsewhere [2,3], the q2DEG at the $LaAlO_3/SrTiO_3$ interface can be effectively tuned by scanning $LaAlO_3$ surface using a biased CAFM probe. In the present work the samples B & C were treated by CAFM under a bias of -8 V, at ambient environment with a relative humidity of ~35 %. A multimode Digital Instruments NANOSCOPE 3100 AFM system and Pt/Ir5 coated silicon tips were used. The entire active device area was scanned with a tip velocity of ~100 µm/s. More details of the experimental setup can be found elsewhere [2].

**4. Electrical measurements.** The samples were measured using a standard Hall bar configuration with a current in a range of 0.2-1 $\mu$A. The electrical contacts to the buried q2DEG were made by ultrasonic bonding with Al wires. The samples A, C & D were measured in a $^4$He cryostat with a $^3$He insert. The sample B was measured in a dilution refrigerator with a base temperature of 10 mK.

**SM B. Supplementary transport data**

As shown in Fig. S1, all of the four samples show metallic conduction behavior down to low temperatures. Figure S2 shows that $R_{xx}$ is rather symmetric with regards to the magnetic field $B$ for sample A. As shown for sample A, the extrema in the $R_{xx}$ and $d^2R_{xx}/dB^2$ data coincide excellently, except for a π shift in phase [Fig. S3]. Figures S4(a) & S4(b) show the temperature dependences of $R_{xx}(B)$ and $R_{xy}(B)$ for sample A.

**SM C. Inverse Fourier transform (IFT)**

We performed the IFT of the two frequencies, $f_{xx1}$ and $f_{xx2}$, of the FT data of $\mathrm{d}R_{xx}^2/\mathrm{d}B^2$ versus $1/B$ plot of the sample A. The result is shown in Fig. S5. It is clear that the oscillations in low field range ($B$ < 4.3 T) corresponds to $f_{xx1}$ and the oscillations in high field range ($B$ > 4.3 T) corresponds to the superposition of $f_{xx1}$ and $f_{xx2}$. The contribution of $f_{xx2}$ dominates the oscillations in the high field range. This result is in close agreement with the fan diagram analysis [Fig. 3(c) in the main text] which gives two different slopes in the low- and high-$B$ ranges. In the fan diagram analysis only the dominant oscillations from $f_{xx2}$ is distinguished in the high field range.

**SM D. Effective mass $m^*$ and Dingle temperature $T_D$**

We estimated $m^*$ and $T_D$ of the q2DEG from the temperature-dependent amplitude of the SdH oscillations. Taking sample A as an example, we used the data as indicated in Fig. S4(a). The non-oscillatory background was removed by performing second order differentiation of $R_{xx}$ with $B$. Figure S6(a) shows the oscillatory part of $R_{xx}$, $\Delta R_{xx}$, of the sample A, as a function of $B$ at various temperatures. The amplitude of the oscillations is defined as half of the difference between neighboring local minima and maxima in the $\Delta R_{xx}(B)$ data, where $B$ is taken as value at the midpoint.

Figures S6(b) & S6(c) show the temperature dependence of amplitude of SdH oscillations at $B$ = 4.1 and 10.8 T, respectively. These data were fitted using the LK formula [4,5]

$$\Delta R_{xx} = 4R_0 \exp(\frac{-4\pi^3 k_B T_D}{h\omega_c})\frac{4\pi^3 k_B T / h\omega_c}{\sinh(4\pi^3 k_B T / h\omega_c)},$$

where $R_0$ is the non-oscillatory component of the sheet resistance, $\omega_c$ is the cyclotron frequency ($eB/m^*$), and $k_B$ is Boltzmann's constant. The best fit of these data give $m^*$ = 0.9 ± 0.1 and 1.30 ± 0.04 $m_e$, $T_D$ = 2.0 ± 0.2 and 1.27 ± 0.04 K, respectively, for $B$ = 4.1 and 10.8 T. The error bars in $m^*$ and $T_D$ only reflect the errors from the fitting process. Similar estimations were performed on other samples and the results are summarized in Table 1 in the main text.

**SM E. Apparent spin degeneracy**

The resolution of the SdH oscillations of the present q2DEG is estimated to be ~ 0.2 meV, based on the fact that the SdH oscillations are essentially undetectable for $B$ < 2 T, and the separation of Landau levels, $heB/2\pi m^*$, is ~ 0.2 meV at $B$ = 2 T. As discussed in the main text, the Zeeman spin splitting is comparable to the Landau splitting, and thus the intrinsic spin degeneracy of each Landau level should always be lifted. However, as sketched in Fig. S7(a), the spin-split up (down) levels may overlap with the adjacent down (up) levels when the splitting energies of the Zeeman spin splitting and the Landau splitting are very close. Experimentally we cannot resolve the spin-split Landau levels if their separation is less than 0.2 meV, causing an effective apparent spin degeneracy. Given $m^* = 1.3\ m_e$, a simple estimation shows that the spin-degeneracy can be resolved at $B > 10$ T when $1.9 < g < 2.7$. As a demonstration, in Figs. S7(b) – 7(d) we simulated the Landau level spectrum using $E_{N\pm} = \frac{h}{2\pi}\omega_c(\mathrm{N}+\frac{1}{2}) \pm \frac{1}{2} g\mu_B B$, where $N$ = 0, 1, 2…etc. We assume $m^*$=1.3 $m_e$, and choose $g$ = 1.6, 2, and 2.4. The apparent spin degeneracy is evident in the case of $g$ = 1.6.

**SM F. Cyclotron effective mass of heavy electrons**

The geometrical definition of the cyclotron effective mass [6,7], $m^*$, is $m^* = \frac{h^2}{8\pi^3}\frac{\partial A}{\partial E}|_{E=E_F}$, where $A$ is the enclosed area of the electronic orbit in momentum space. From this definition and Fig. 4(a) in the main text, we have $m_l^* = \frac{h^2}{8\pi^3}\frac{\partial A(\mathrm{circle})}{\partial E}|_{E=E_F}$ ,and $m_h^* = \frac{h^2}{8\pi^3}\frac{\partial A(star)}{\partial E}|_{E=E_F}$ for the light and heavy electrons, respectively. For the magnetic breakdown orbit MB1, we have $m^*(MB1) = \frac{h^2}{8\pi^3}\frac{\partial A(MB1)}{\partial E}|_{E=E_F} \cong \frac{h^2}{8\pi^3}\frac{\partial[\frac{1}{4}A(star)+\frac{3}{4}A(circle)]}{\partial E}|_{E=E_F} = \frac{1}{4}m_h^* + \frac{3}{4}m_l^*$. Similarly, for the magnetic breakdown orbit MB2, we have $m^*(MB2) \cong \frac{h^2}{8\pi^3}\frac{\partial[\frac{1}{2}A(star)+\frac{1}{2}A(circle)]}{\partial E}|_{E=E_F} = \frac{1}{2}m_h^* + \frac{1}{2}m_l^*$. For the case a) discussed in the main text, using $m^*(MB1) = 1.30 \pm 0.04 m_e$ and $m_l^* = 0.9 \pm 0.1 m_e$, we obtain $m_h^*$ = 2.5 ± 0.5 $m_e$. For the case b), using $m^*(MB2) = 1.30 \pm 0.04 m_e$ and $m_l^* = 0.9 \pm 0.1 m_e$, we find $m_h^*$ = 1.7 ± 0.2 $m_e$.

**SM G. Fermi energy and the enclosed area of magnetic breakdown orbit**

In the scenario depicted in Fig. 4 in the main text, the sheet carrier density of light electrons, $n_l$, is equal to $eS_L/h$ or $2eS_L/h$ when spin degeneracy is lifted or kept respectively; the sheet carrier density of heavy electrons, $n_h$, is $n_{Hall} - n_l$. Assuming parabolic dispersion relation, the bottom of the light band is below the Fermi level at $E_L = \frac{heS_L}{2\pi m_l^*}$ = 2.5 meV; the bottom of the heavy band is below the Fermi level at $E_F = \frac{h^2 n_h}{2\pi m_h^*}$ (when spin degeneracy is lifted) or $E_F = \frac{h^2 n_h}{4\pi m_h^*}$ (with effective spin degeneracy of 2). From these relations we obtain $E_F$ = 8.0 meV and 5.2 meV for the cases a) and b), respectively.

From the geometrical relation between the sheet carrier density and the area of the Fermi surface, $A_F = 4\pi^2 \times n_{2D}$/degeneracy, in the case a) we have

$$A(star) = 4\pi^2 n_h = \frac{n_h}{n_l} \times 4\pi^2 n_l = \frac{n_h}{n_l} A(circle) = 8.8 A(circle).$$

Hence $A(MB1) = \frac{1}{4} A(star) + \frac{3}{4} A(circle) = 2.95 A(circle).$ Similarly, in the case b) we have $A(star) = 3.8 A(circle)$, and $A(MB2) = \frac{1}{2} A(star) + \frac{1}{2} A(circle) = 2.4 A(circle)$.

**SM H. Rashba spin-orbit coupling**

Rashba spin-orbit coupling is expected to be present to some degree due to the lack of inversion symmetry at the $LaAlO_3/SrTiO_3$ interface, although there is a lack of consensus currently on its microscopic form and magnitude [8–13]. Figure S8 shows a schematic diagram of Rashba spin split subband structure, assuming linear spin-orbit coupling. In this scenario both $S_H/S_L > 2$ and $m^*$(high-$B$) > $m^*$(low-$B$) are natural consequences if only the light band is occupied, in good analogy with what has been observed in other 2D systems with Rashba spin-orbit coupling [14–16]. $S_H$ can be assigned to the total density of oscillating electrons [14–16]. Using the Onsager relation, we extract $n_s = eS_H/h = 1.45 \times 10^{12}$ cm$^{-2}$ (ignoring any degeneracy). Taking into account the 4-fold degeneracy indicated from the plateaus in $G_{xy}(B)$, we find $n_s = 5.8 \times 10^{12}$ cm$^{-2}$, agreeing reasonably with $n_{Hall}$ (Table 1 in the main text).

The Rashba constant $\alpha$ can be calculated from [17,18] $\sim = \frac{h^2}{8\text{®}^2 m^*}(k_{F+} - k_{F-})$ , where $k_{F\pm} = (4\text{®} N_\pm)^{\frac{1}{2}}$. Following the analyses used in the traditional 2D systems with Rashba spin-orbit coupling [14,15], we assign $f_{xx1}$ to the inside Fermi surface, and $f_{xx2}$ to the sum of the inside and outside Fermi surface. Hence, we have $N_- = ef_{xx1}/h$ and $N_+ = e(f_{xx2} - f_{xx1})/h$. Using the above relations and with $m^* = 1.3\ m_e$, $\alpha$ is calculated to be $2.93 \times 10^{-12}$ eVm, $2.6 \times 10^{-12}$ eVm, $2.6 \times 10^{-12}$ eVm, and $1.1 \times 10^{-12}$ eVm for the samples A, B, C, and D, respectively. These values are reasonably consistent with those deduced from a weak-localization study [8].

One scenario to explain the degeneracy is that only the light band ($3d_{xy}$) derived subbands are occupied. The interfacial quantum confinement effect is weak due to the large out-of-plane effective mass [19] and the very low $n_s$ [20]. As a consequence, the separation of the quantum well states is small and there exist a few closely distributed subbands below $E_F$. The Landau levels of these subbands may overlap and smear each other, causing the apparent degeneracy. However, if the Rashba scenario is correct, we require an accidental 4-fold degeneracy to explain the present observations, which is unlikely. In addition, Rashba spin-orbit coupling in the present samples are expected to be weak because $n_s$ is low and thus the electric field across the q2DEG should be weak. Previous energy band calculation [21] has suggested that in $LaAlO_3/SrTiO_3$ heterostructures atomic spin-orbit coupling is much stronger than the Rashba spin-orbit coupling. Rather, it is likely that spin-orbit coupling plays an important role for strong magnetic breakdown effects, since even a relatively small spin-orbit coupling modifies the band structure most significantly close to subband crossings.

## Supplementary References

## Supplementary Figures and Legends

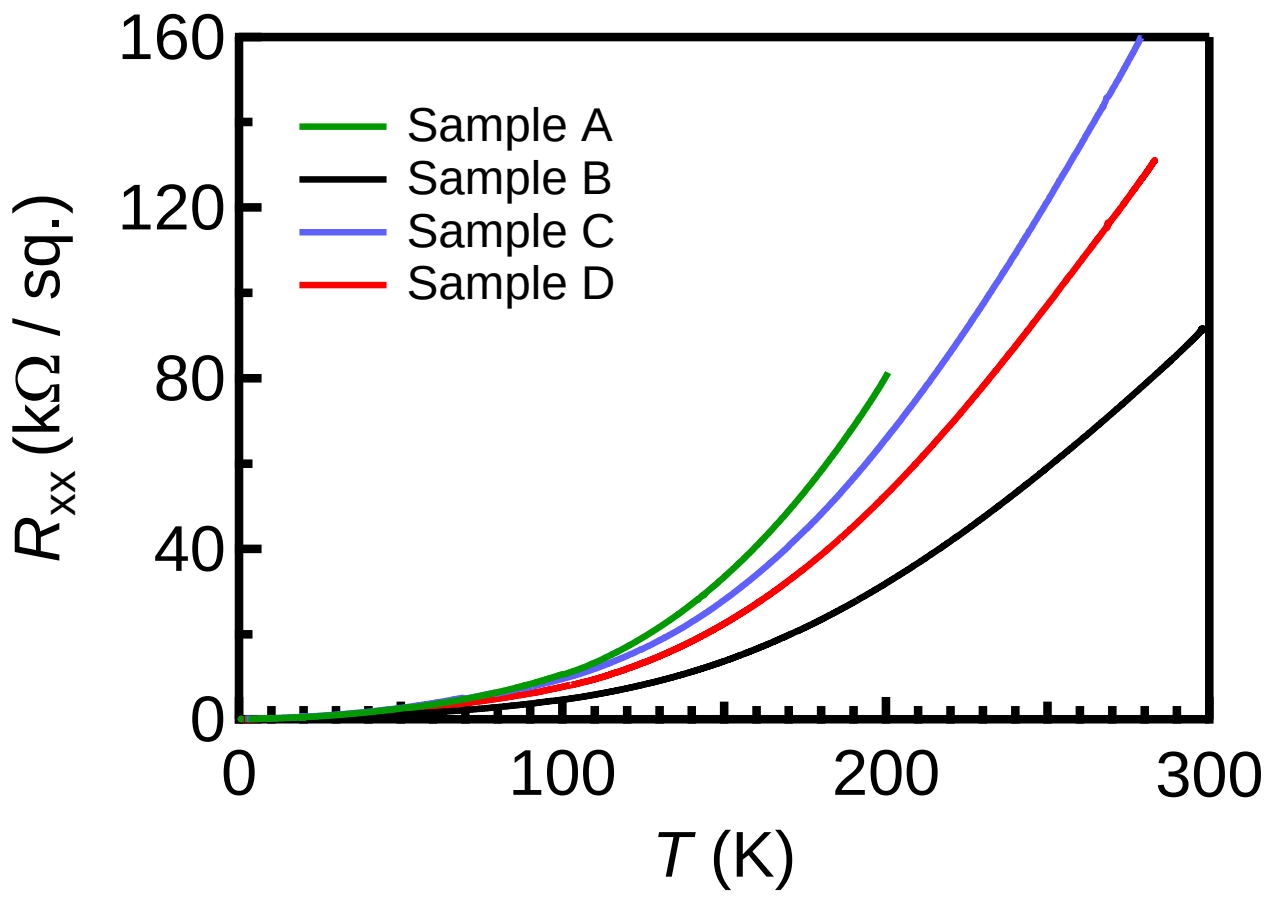

**Fig. S1.** Temperature dependence of sheet resistance, $R_{xx}$, of the four samples.

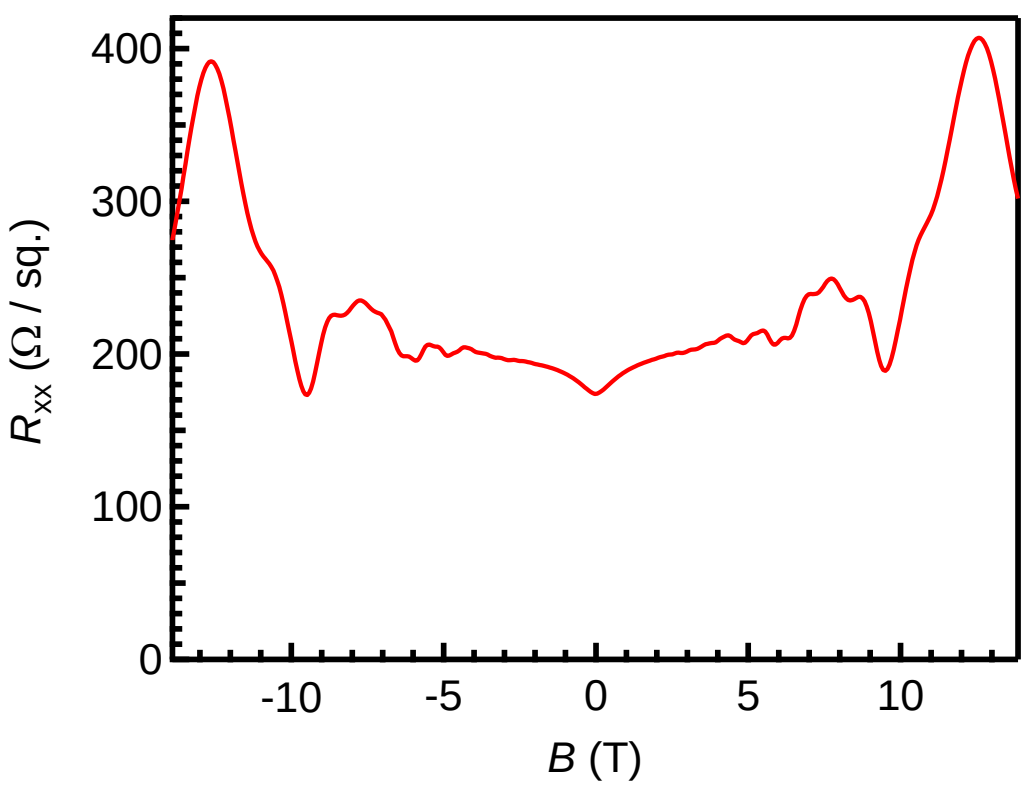

**Fig. S2.** Raw $R_{xx}$ of sample A as a function of magnetic field $B$ from -14 to 14 T, measured at $T$ = 0.35 K.

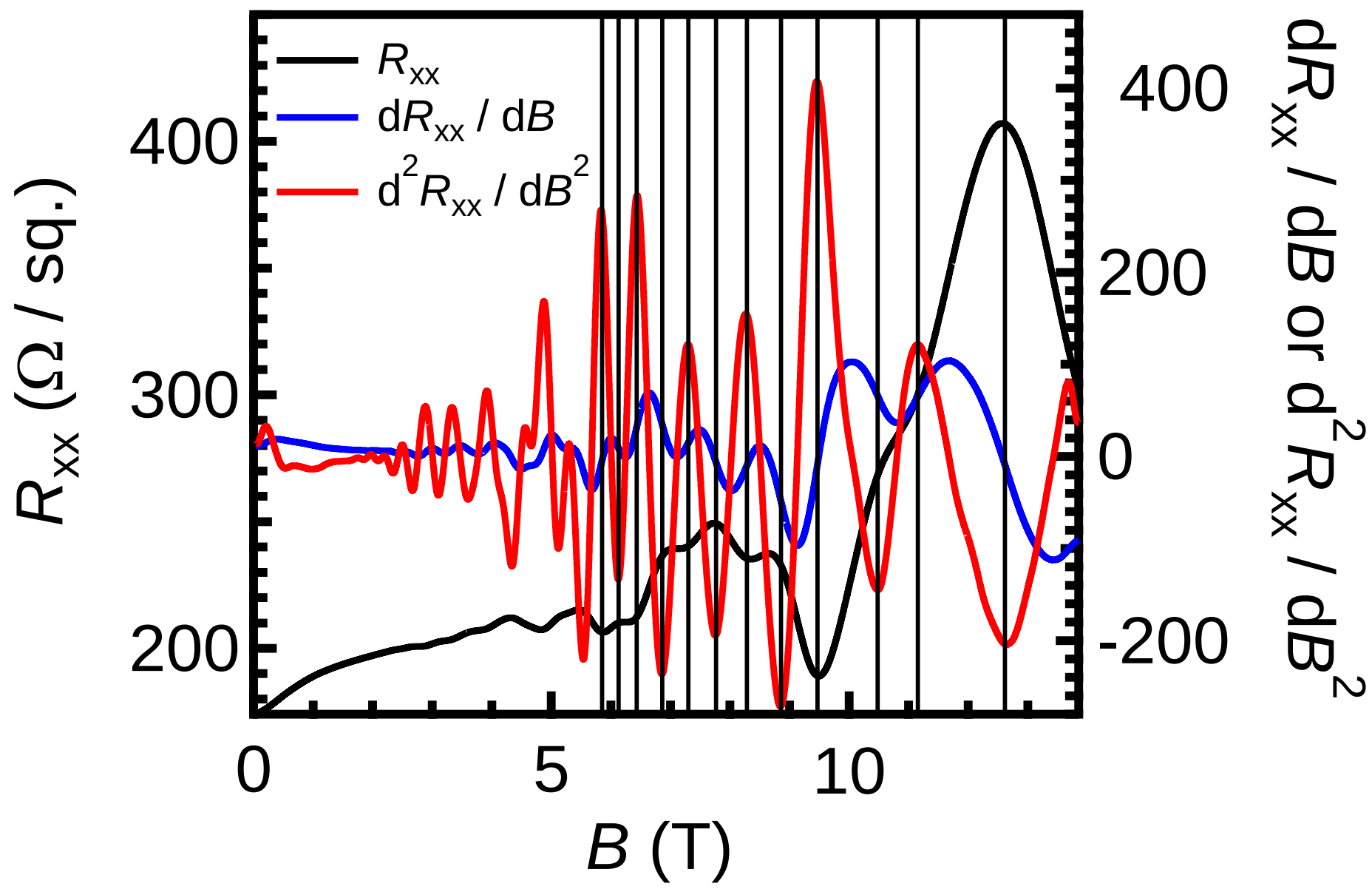

**Fig. S3.** $R_{xx}$ and its first derivative, $dR_{xx}/dB$, and second derivative, $d^2R_{xx}/dB^2$, of sample A as a function of magnetic field $B$. The extrema in $R_{xx}$ and $d^2R_{xx}/dB^2$ coincide excellently except for a π shift in phase (maximum vs. minimum). The dashed lines are guides for the eye.

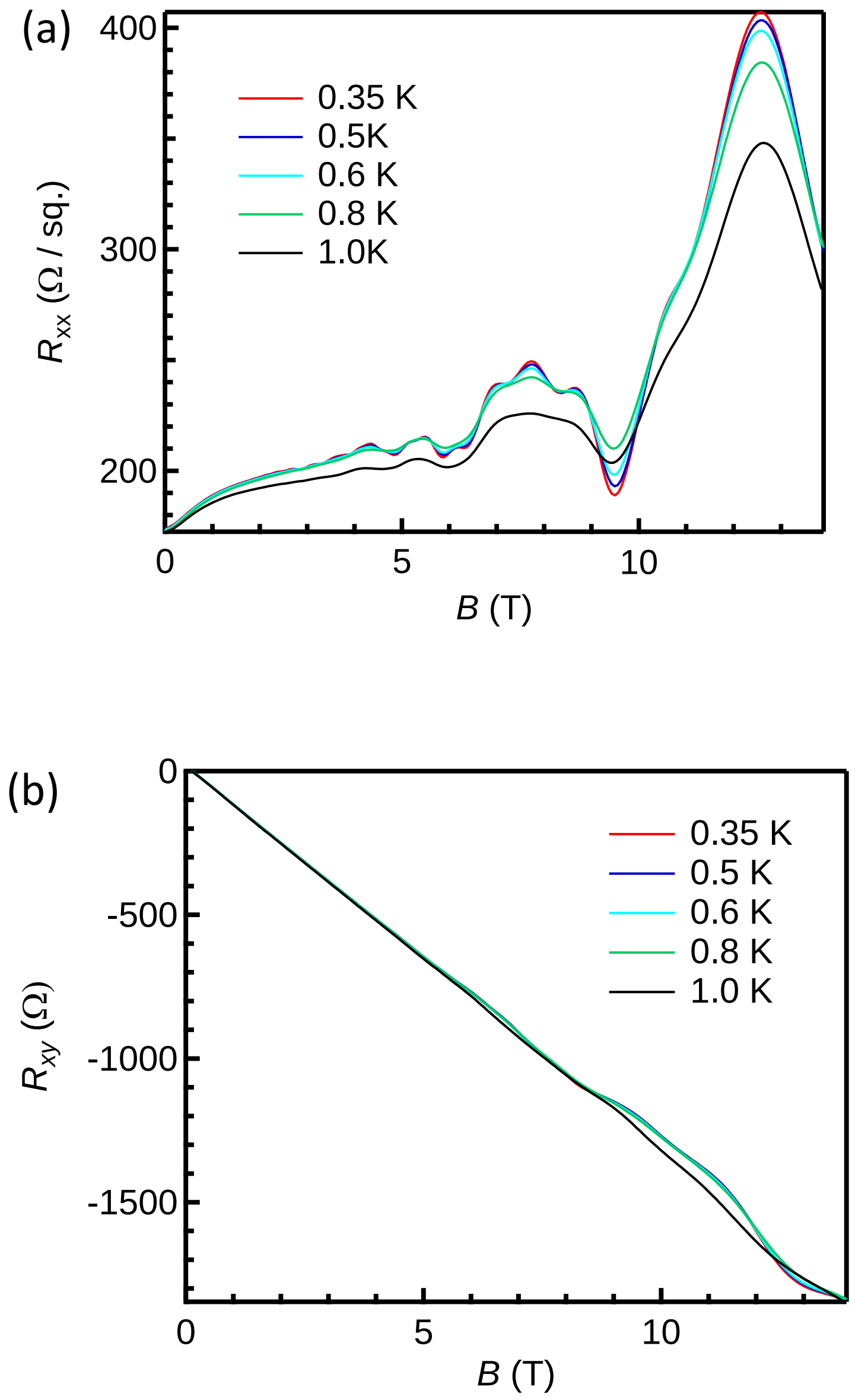

**Fig. S4.** Temperature dependence of (a) $R_{xx}$ and (b) $R_{xy}$ of sample A as a function of magnetic field *B*.

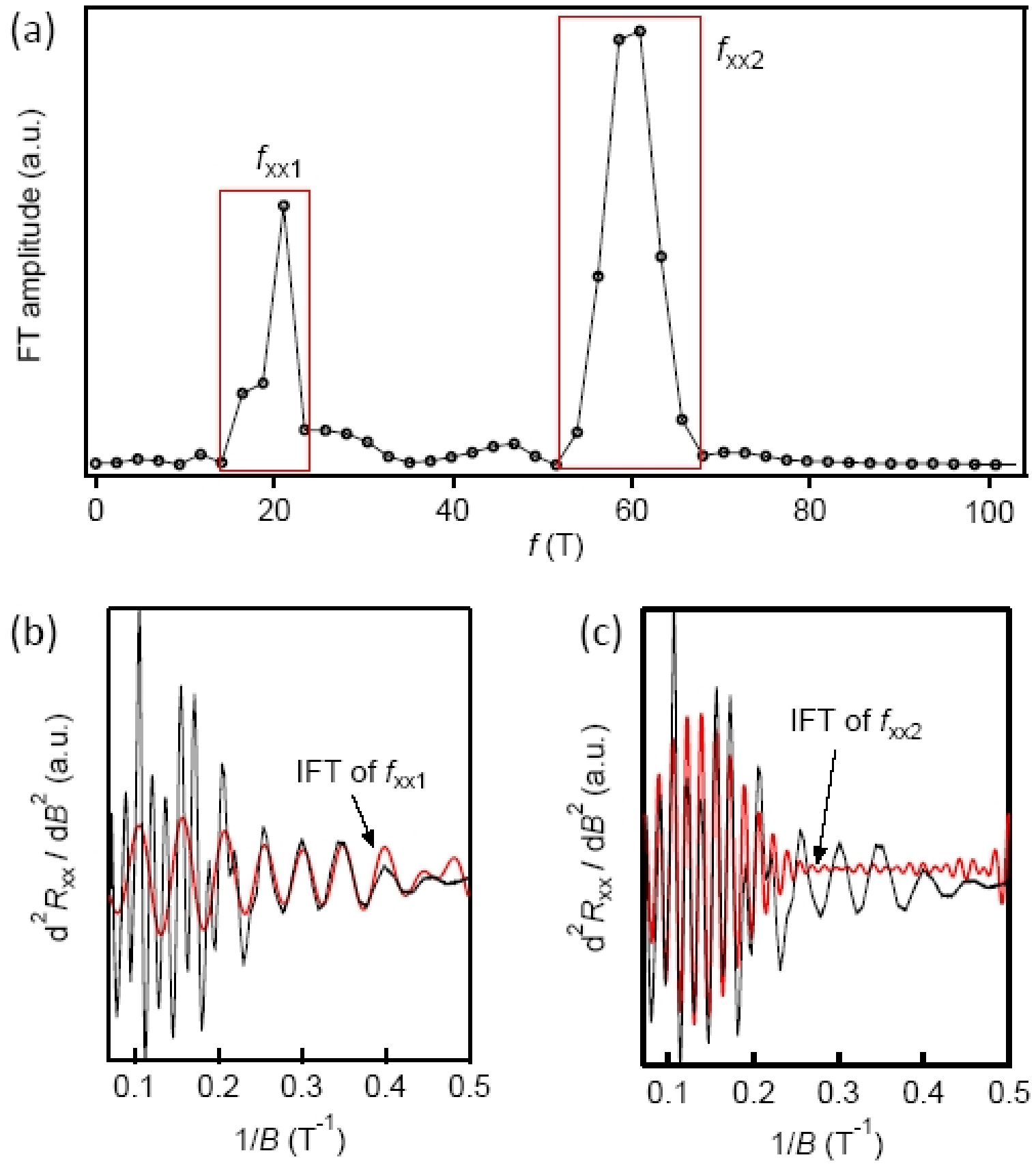

**Fig. S5.** Inverse Fourier transform (IFT) of the FT data of $dR_{xx}^2/dB^2$ of sample A. (a) FT of $dR_{xx}^2/dB^2$ versus $1/B$ plot. (b) IFT of the boxed area enclosing $f_{xx1}$ in (a) (red plot). (c) IFT of the boxed area enclosing $f_{xx2}$ in (a) (red plot). The $dR_{xx}^2/dB^2$ versus $1/B$ is shown as the black plots in (b) and (c).

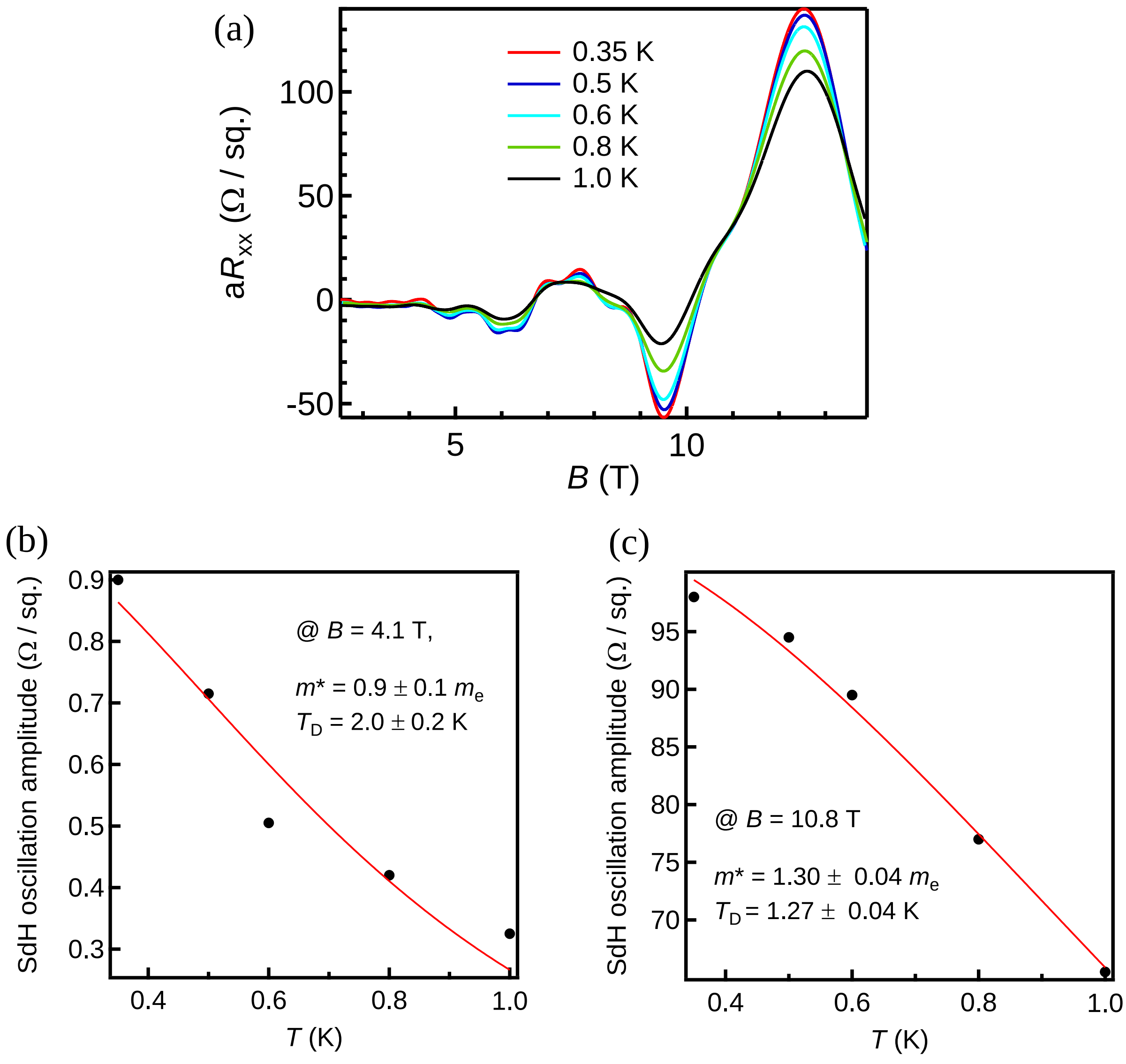

**Fig. S6.** (a) The oscillatory part of $R_{xx}$, $\Delta R_{xx}$, of sample A, as a function of $B$ at various temperatures. The temperature dependence of the amplitude (dot symbols) of the SdH oscillations obtained at (b) $B$ = 4.1 T and (c) 10.8 T. Here the amplitude was defined as described in the text in Section D. The solid curves in (b) and (c) are the best fits to the data as described in the text.

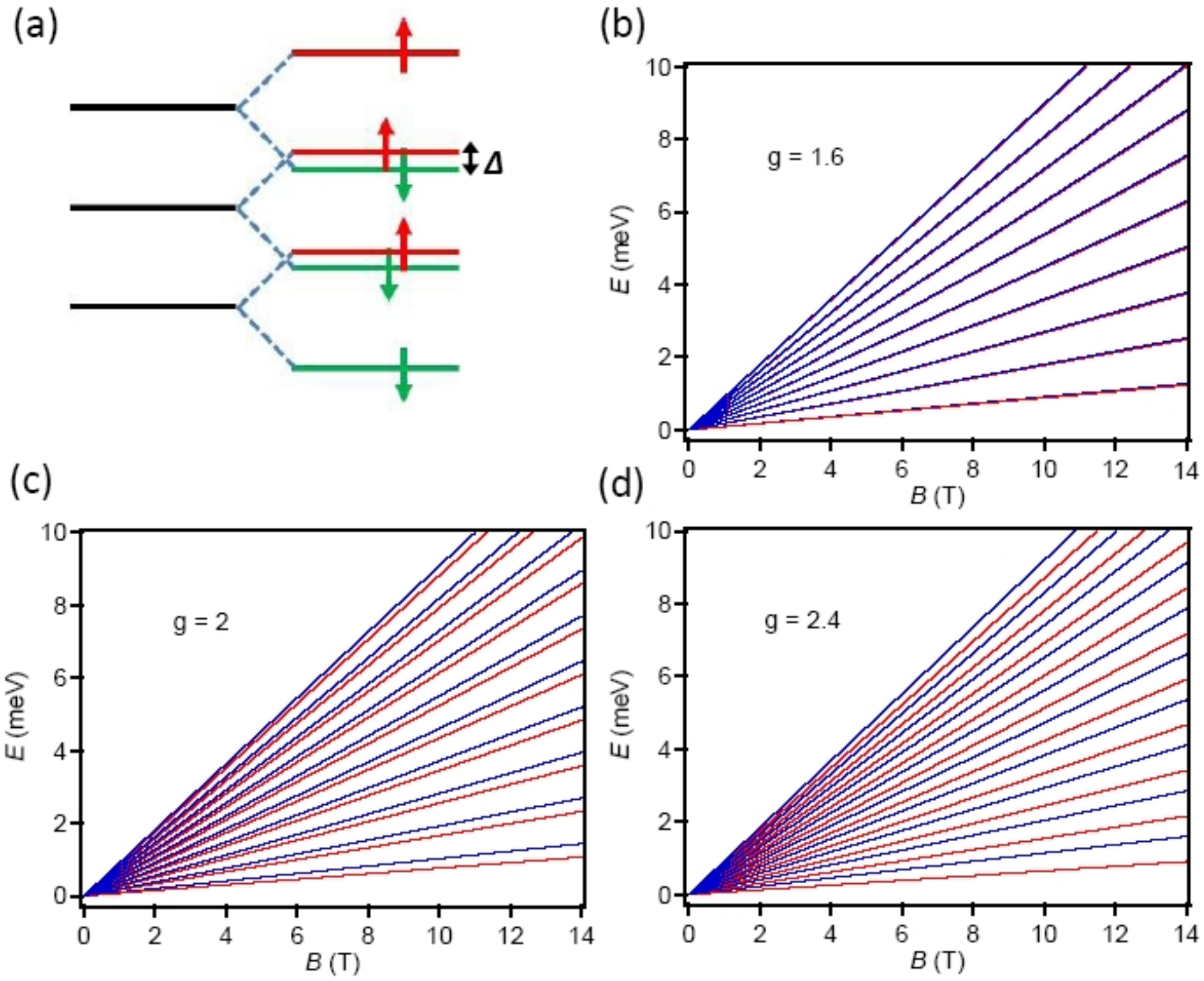

**Fig. S7.** (a) Landau levels without (left) and with (right) Zeeman spin splitting. There is apparent spin degeneracy when $\Delta$ is small. Simulated Landau level spectrum with $B$, assuming $m^* = 1.3\ m_e$, and (b) $g = 1.6$, (c) $g = 2$, and (d) $g = 2.4$. The red lines are $E_{N+}$ branch and the blue lines are $E_{N-}$ branch. For each branch the index of levels increases from bottom to top from 0 to 1, 2, 3… etc.

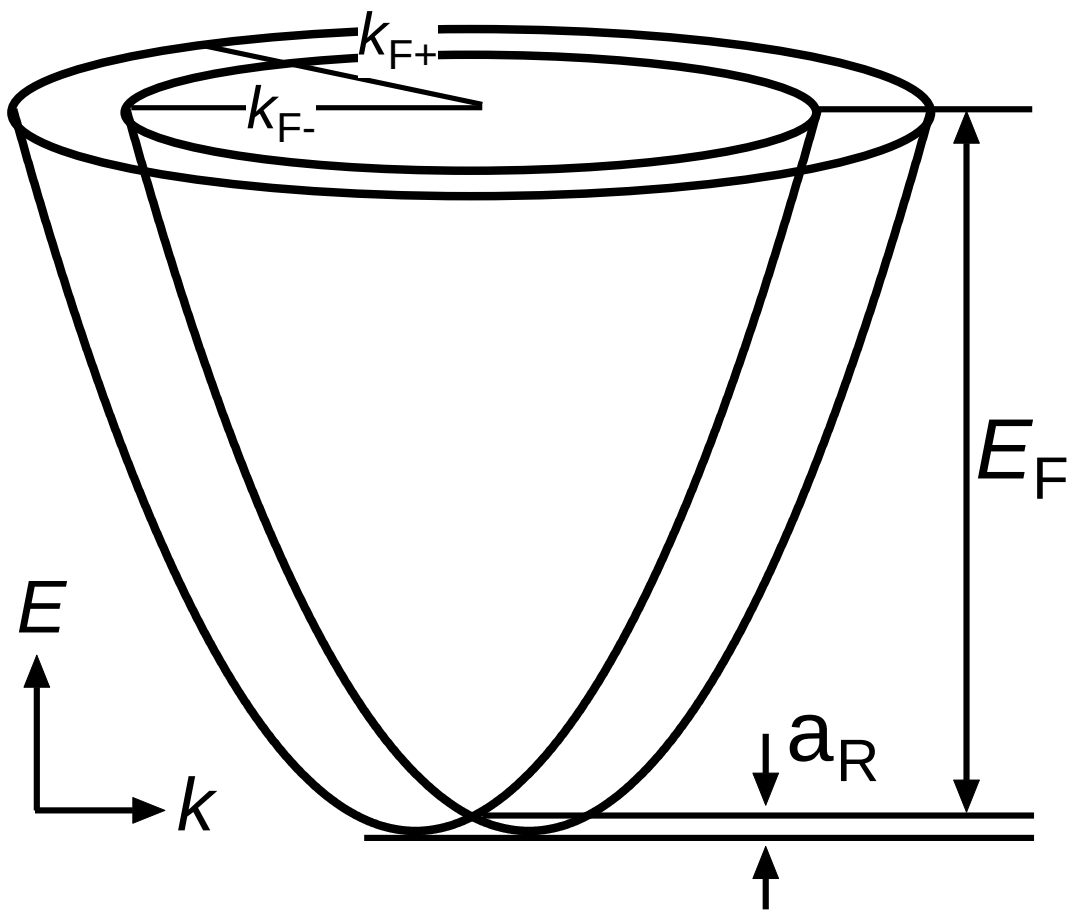

**Fig. S8.** A schematic diagram of single-band Rashba spin split subband structure.